# New views of old proteins:
# clarifying the enigmatic proteome

### *Participants in a NIH Workshop on Functional and Integrative Proteomics*

Kristin E. Burnum-Johnson[1*], Thomas P. Conrads[2], Richard R. Drake[3], Amy E. Herr[4], Ravi Iyengar[5], Ryan T. Kelly[6], Emma Lundberg[7], Michael J. MacCoss[8], Alexandra Naba[9], Garry P. Nolan[10], Pavel A. Pevzner[11], Karin D. Rodland[12], Salvatore Sechi[13], Nikolai Slavov[14], Jeffrey M. Spraggins[15], Jennifer E. Van Eyk[16], Marc Vidal[17], Christine Vogel[18], David R. Walt[19] and Neil L. Kelleher[20*]

*Abstract.* All human diseases involve proteins, yet our current tools to characterize and quantify them are limited. To better elucidate proteins across space, time, and molecular composition, we provide provocative projections for technologies to meet the challenges that protein biology presents. With a broad perspective, we discuss grand opportunities to transition the science of proteomics into a more propulsive enterprise. Extrapolating recent trends, we offer potential futures for a next generation of disruptive approaches to define, quantify and visualize the multiple dimensions of the proteome, thereby transforming our understanding and interactions with human disease in the coming decade.

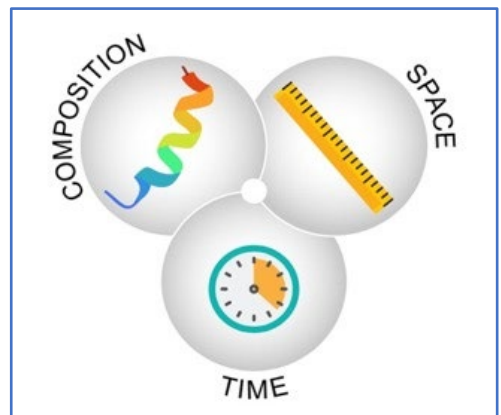

**TOC Image.** Capturing the biology of proteins will require improved technologies to readout their composition in space and time. Developing these improved technologies presents a major opportunity for biomedical research. How might we proceed in the decade ahead?

Table of Contents

## Section 1. The proteome is highly complex

Proteins are the primary conduit connecting our genes to complex traits including the diseases that afflict us. While the genome provides the 'biological script,' proteins run and fill it with life. Given this critical role, the enormous complexity of the proteome, and the pace of technology development, we highlight strategies and opportunities for proteomics in the coming decade. Compared to the 30–40× coverage of DNA in the genome, our ability to detect or sequence a proteome offers only ~0.1× coverage. This proteomic gap must be addressed in the coming years.[1] We therefore argue for accelerated technology development, concerted efforts across laboratories, and large consortia with meaningful data integration. Needed advances in measurement science will match the complexity of our biology so new enterprises emerge to boost detection and interventions that will reliably extend our life spans and quality of life.

The immense complexity of the proteome arises from the fact that each gene produces several protein variants far beyond what can be predicted from DNA sequence. The term 'proteoform' captures this molecular complexity (**Figure 1**). The proteoform paradigm describes the composition of protein molecules arising from ~250 types of posttranslational modifications (PTMs),[2] alternative transcriptions or translation start sites, or amino acid changes in our population (**Figure 1**). A single cell easily encompasses hundreds of thousands of different proteoforms reflecting the true chemical diversity of protein molecules,[3,4] each having the potential to alter structure, function, and interactions. While the number of theoretically possible proteoforms is extremely large,[3,4] existing methods detect fewer than 10,000 proteoforms in cell lines today.

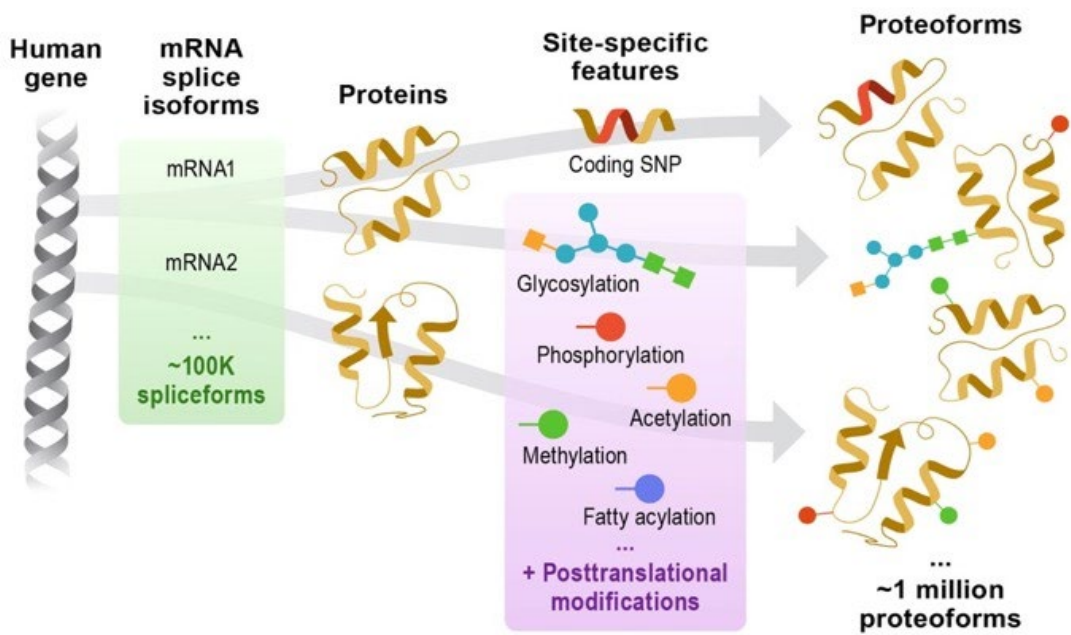

**Figure 1.** From human genes (far left), diverse forms of mature, endogenous protein molecules are expressed (far right). Technologies are needed to fully measure the dynamic set of ~1 million distinct proteoforms expressed in each cell (see Aebersold *et al.*, 2018).

Given the complexity of proteins, it is unsurprising that RNA abundance or ribosome-bound mRNA are imperfect surrogates for protein abundance. RNA abundance ranges over ~4 orders of magnitude in cells, while proteoforms range over ~7 orders of magnitude and even over 10 in body fluids.[5] Researchers use diverse proteomic technologies to identify and quantify proteins, and the analytical sensitivity, reproducibility, and specificity of platforms is extensive. These tools enable both discovery experiments (where emphasis is on identifying large numbers of proteins) and targeted measurements that focus on individual proteins, pathways, and/or certain classes such as membrane proteins. Ongoing innovation in spatial proteomics and multiplexing of compositional proteomics continue to improve our ability to probe the approximately half of the 20,000 genes expressed into detectable proteins in a given cell type. Modern proteomics has advanced our functional and mechanistic insights of proteins incredibly over the past 25 years[6].

Looking forward, one significant goal in the coming decade is to increase the efficiency of and access to different proteomic platforms while also decreasing cost. We need to prioritize technology development to deliver diverse data types of high value to the community. Efforts are underway; for example, mapping projects that use multiplexed protein measurements with single cell resolution via antibody-based tissue imaging. The Human Protein Atlas[7] and the Human BioMolecular Atlas Program (HuBMAP)[8] are examples of such efforts. But, if functional proteomics aims to capture biological complexity in its dynamics and relationships across scales, we need to upgrade our ability to spatially sample proteoforms, their complexes in organelles, and in regions of phase transition to make functional assertions with high confidence. This will mean creating technologies capable of molecular deconvolution from millimeters to nanometers and their temporal dynamics from seconds to years.

Another significant goal will rely on integration of multi-omic data with spatial proteoform data. Accomplishing an upgrade for spatial multi-omics will allow us to more reliably infer emergent properties of whole organisms. Integration is still challenging due to missing values and the fact that defining proteins precisely is challenging, even at the level of their full sequence and composition. Still, these two goals–new technology combined with meaningful data integration–are achievable within the next 10 years.

## Section 2. PTMs are major drivers of proteoform diversity

### 2a. Top-down proteomics

A first step toward comprehensively understanding proteome complexity is the efficient detection of proteoforms and their PTMs. PTMs are one of the major routes to produce proteoforms and many proteins have multiple modifications, but it is still unclear which of these PTMs are functional and which are passengers or "bystanders".

With "top-down" mass spectrometry (MS), intact proteoforms and their assemblies are analyzed, thereby directly informing on co-occurring events (i.e., PTMs, isoforms, or mutations). This approach has some advantages over peptide-centric, "bottom-up" approaches described below, including:[9] resolving the protein inference problem[10], identifying isoforms directly, and retaining both stoichiometry and the combinatorial occurrence of PTMs (**Figure 2**)[9]. Because so many questions about PTMs remain unknown, expanding this approach will improve the efficiency of detection and assignment of function to proteoforms and their PTMs. Proteoform knowledge will also accelerate the development of disruptive technologies to improve biomarker discovery in translational proteomics by more accurately annotating the chemical diversity of protein variation in disease models and patient cohorts.

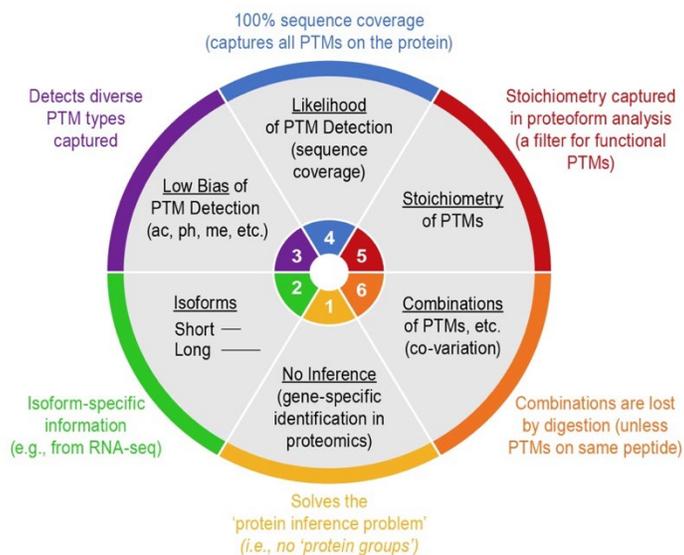

**Figure 2.** *Proteoform measurement fills knowledge gaps by measuring intact proteins.* Measuring proteoforms will improve molecular precision in asserting protein sequence and composition (i.e., primary structure). Such measurements should improve the efficiency of detection and assignment of function to diverse PTMs in the coming decade.

## 2b. Bottom-up proteomics

In detecting thousands of peptides and PTMs efficiently, bottom-up proteomics uses proteases to digest the proteome and is the current 'go-to' platform for discovery proteomics. Unlike whole proteins, many peptides are easily fractionated, ionized, and fragmented via liquid chromatography-tandem MS methods. In addition, several options exist for collecting information on peptides and modified peptides and for infusing spatial information (**Figure 3**, **upper right**).[11] However, data from bottom-up approaches remain limited in sequence coverage of individual proteins and coverage of the single-cell proteome. For example, biological signals are masked by combining peptides into a single protein quantity, losing information on combinatorial effects.[12] Targeted MS assays, such as selected reaction monitoring (SRM) can precisely identify peptide sequences and specific PTMs to achieve absolute quantification and probe known proteoforms.[13,14] Future developments should continue by combining top-down with bottom-up strategies in an integrated approach for robust quantitation of isoforms and PTMs.

## Section 3.  Proteomics goes single cell (and single molecule)

### 3a. Single-cell proteomics

Tissues are heterogeneous mosaics of cells in different physiological stages, so single-cell analysis is required. However, given that we cannot currently amplify proteins and their concentrations cover a large dynamic range, detecting them at the single-cell level is challenging.

Some single-cell proteomics methods employing antibodies or other affinity reagents have existed for decades (**Figure 3, left**). While these methods have improved significantly in power and sophistication,[15] the newest developments via bottom-up mass spectrometry have clearly expanded the horizon for single-cell proteomics.[16-27]

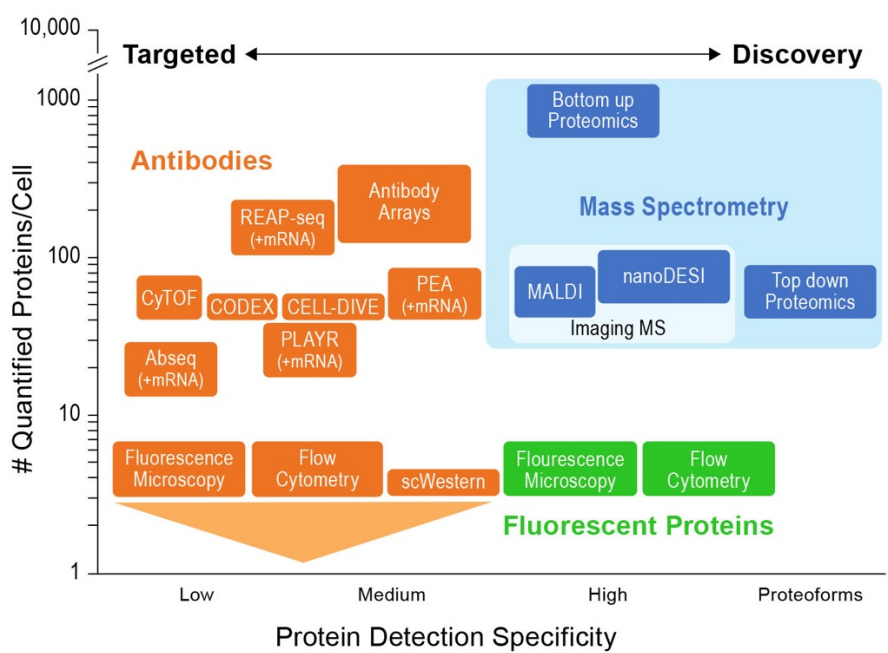

**Figure 3.** *Multiple approaches for assessment of proteins in single cells.* (*Legend:* CyTOF: mass cytometry; CODEX: co-detection by antibody indexing; sc: single cell; Top-down MS: measures proteoforms; Bottom-up MS: measures peptides; nanoDESI: spatially resolved desorption by electrospray; REAP: a rapid, efficient and practical cell processing method; PEA: proximity extension assay; PLAYR: proximity ligation assay for RNA.)

Over the last four years, mass spectrometric methods have advanced from quantifying a few hundred proteins per cell to quantifying more than 1000 proteins per cell, with thousands of proteins quantified across data sets at a throughput of more than 200 single cells per day (**Figure 3, upper right**).[28,29] One enabling technology making such in-depth single-cell and 'low-input' proteome profiling possible is multiplexed analysis of single cells labeled with isobaric mass-tags and combined with an isobaric carrier[26] to enhance throughput and MS/MS peptide sequence identification.[30] Miniaturization of sample preparation volumes to nanoliter or low-microliter levels has minimized surface losses and enhanced reaction kinetics.[31,32] In addition, miniaturization of liquid chromatography[25] and capillary electrophoresis[33] separations have boosted prospects, even adding peptide selectivity to the workflow.[24,27] However, future approaches still need to achieve higher coverage and throughput more comparable to those of single-cell RNA-seq. We posit that this will be achievable given the rate of improvement. Integration of proteomics data with other single-cell modalities, such as scRNA-seq, have already provided initial results.[16,20,28] We imagine that within the next 5 years, proteomics will cover >5000 proteins from thousands of single cells per day, with linked information on transcripts.

## 3b. Single molecule sequencing

Another promising area of technology development includes single-molecule protein sequencing and proteoform detection.[34-37] These approaches have the potential to be disruptive and to impact single-cell protein analysis, with sequencing demonstrated only on synthetic peptides thus far. Intriguingly, there have been major increases in private sector investments in this general area over the past year, creating a palpable sense that this could be the decade of breakthroughs in single molecule protein measurement and sequencing.

Such disruptive, scalable technologies operative on single molecules will encounter the challenge of proteoform diversity expressed in human biology (**Figure 1**). Therefore, they would benefit from better compositional definition of the expressed proteome. Indeed, amassing a reference set of human proteoforms has been proposed[6,38,39] using both a cell- and gene-based approaches.[38] Using output of the Human Cell Atlas, HuBMAP and other consortia, the number of human cell types will be determined, allowing the proteome of each type to be targeted and better defined. Assuming, 5000 cell types and 1 million proteoforms in each cell type, such a human proteoform atlas would involve approximately 5 billion measurements of redundant proteoforms, with perhaps approximately 50 million unique proteoforms needing to be asserted and mapped in the coming decade.[38] As they ratchet up their platforms, disruptive proteomics technologies would benefit from atlasing efforts, providing reference sets of what proteoforms are actually present and where.

## 3c. Single cell and single molecule immunoassays

Complementing methods for proteome-wide characterization, *targeted assays* using affinity reagents can accurately detect and quantify specific proteins. While often targeting fewer proteins, they achieve far higher throughput than non-targeted approaches. Furthermore, many proteins of interest in the human proteome are below the limit of detection of current MS methods and conventional immunoassays. One major group of targeted approaches includes immunoassays that are designed to detect known protein targets, with target identities informed by *a priori* discovery proteomics or driven by biological hypotheses. At the single-cell level, the formats are diverse and include immunohistochemistry, flow cytometry, mass cytometry and various formats for multiplexed antibody-based imaging. Arrays of affinity reagents are also available for limited operation in discovery-mode; use of these platforms is likely to increase in the coming decade.

Over the last two decades, miniaturization of single-cell immunoassays has improved to the point where proteins in thousands of samples can now be processed in parallel, while using far less time, sample, and reagents. Performance improvements stem from the underlying physics at miniature scale and new assay designs made possible by microfluidic approaches.[40] Microdevices afford scalable means to precisely compartmentalize individual cells, with the use of both droplets[41] and microwells being common strategies.[42] Further, ultrasensitive single-molecule technologies already access proteins at picomolar and femtomolar concentrations in the single-protein regime using digital immunoassays like SIMPul,[43] Simoa,[44] and droplet-based assays.[45,46] While not providing proteoform-level specificity, these single-cell frameworks are powerful for multi-modal, same-cell measurements, with an immunoassay as one of the reported modes.[47,48] Scrutinizing the content of targeted proteins in lysates from thousands of individual organelles is viable today using microfluidic devices to overcome shortcomings of single capillary systems for single-cell,[49,50] sub-cellular,[51] and multi-modal[52,53] immunoblotting.

Given new developments in droplet microfluidics, microfluidic imaging cytometry,[41] and subcellular immunoblotting, we envision targeted single-cell proteomics to push two new frontiers: 1) monitoring protein expression in specific cellular compartments and 2) multi-modal single-cell analyses that integrate proteomics with other omics data. While selective target detection and antibody availability remain an inherent limitation of immunoassays, we foresee a tremendous opportunity for immunoassays to become proteoform-specific across diverse cells. We posit that clever immunoassay strategies that enhance analytical specificity[54,55] will help translate proteoform knowledge into single-molecule, scalable assays. We project that by 2030, innovations in this area will move such analyses to cost-effective benchtop instruments usable in diverse settings. Analysis of thousands of individual cells per assay, with specificity for hundreds of currently undiscernible proteoforms within each cell is possible. Absolute, not simply relative, quantification of each proteoform beckons.

### 3d. Availability and validity of affinity reagents

Many assays in basic and clinical research rely on antibodies. However, there are proposed[56], but no widely accepted guidelines to determine the validity of these reagents. Furthermore, many recent publications have highlighted that commercial antibodies can fail to detect the intended target in diverse sample types.[57,58] The utility in biological and biomedical research of affinity reagents for characterizing and quantifying proteins and peptides is central; however, it also should elevate the challenge in validating and reproducing many of the antibody-based assays published each year. For this purpose, increasing effort will be required to improve the rigor and reproducibility of antibody-based assays.[59] For example, efforts such as the antibody registry to provide a specific and unique identifier for reagents used.[60]

The proteomics field would greatly benefit from an 'open source' antibody database and library. Antibodies are expensive and reducing their cost through a public database would allow researchers to devote more of their funding to other materials and maximize findings. To enhance reproducibility and operational efficiency, recombinant antibodies with known sequences would create a long-lasting and reproducible resource across laboratories.[59] Increasingly, antibody sequencing and characterization using proteomics is on the rise,[61,62] making an open-source library of antibodies even more feasible in the coming decade.

Another important resource for proteomics of the future would be a proteoform-specific library of affinity reagents covering as many products of human genes as possible. Moving forward, we should promote gene-specific probes and create a generation of proteoform-aware, PTM-recognizing affinity reagents (e.g., phosphorylation site specific antibody reagents).[63] Creating such a resource will be a massive undertaking, but limited scope pilots in which antibodies to known proteoforms of key protein-hubs in cellular regulatory networks (c-Myc, p53, NKkB, histones, *etc.*) would be a very valuable resource and important next step. Construction of such a library of proteoform-specific antibodies, with known epitope binding specificity, critically depends on the reproducibility, specificity, and scalability with which multiple antibodies can probe multiple epitopes and PTMs on proteoforms.

## Section 4.  Thinking outside the cell: extracellular matrix, body fluids, extracellular communication

Not only does one need to understand the single cell proteome, it is also necessary to quantify the cellular-environment and context. Quantification and defining extracellular communication and the cellular environment is required to infer tissue and organ function. The extracellular matrix (ECM) contains highly glycosylated and crosslinked proteins that form the scaffold of cellular anchorage and migration.[64,65] The ECM also liberates and sequesters secreted molecules such as growth factors.[66] It plays critical roles in development, health, and disease via mechanical and chemical signal transduction. While recent proteomic methods tailored to biochemically unique ECM proteins have led to greater understanding of the ECM proteomes[67-71] in many tissues and organs, these methods cannot fully capture the broad dynamic range of the ECM from low-abundance growth factors to hyper-abundant collagens.[72] New imaging MS methods using collagenase digestions targeted to the ECM are beginning to address some spatial-temporal aspects.[73,74] To achieve near-complete coverage and single-cell-level profiling of the matrisome and to capture its dynamics, our community will need to 1) expand methods to identify and isolate specific cell and tissue niches (e.g., laser-capture microdissection[75] and imaging MS), 2) develop reagents to enrich for ECM proteins or classes of ECM proteins (e.g., high-affinity antibodies against ECM-specific PTMs or splice variants), 3) use ECM-specific proteases such as elastase and collagenases instead of trypsin, and 4) apply adequate data analysis workflows.[76,77]

Beyond the ECM, the extracellular milieu contains protein-rich fluids[78-80] (e.g., tissue interstitial fluids, serum/plasma, urine, saliva, etc.) that present unique challenges when it comes to proteomic profiling. Extracellular vesicles and their protein contents are key mediators of cell communications and regulators of cellular functions,[81,82] including those that can affect "in trans" chemoresistance and angiogenesis in cancer.[83] An important activity for advancing functional proteomics will be to convert ultra-trace analysis into a straightforward, deep characterization of extracellular proteomes and to correlate the resulting information with their cells of origin.

## Section 5. The future lies in integration across scales

### 5a. Spatial information

Be it through networks or other models, future efforts will have to integrate compositional, interactomics, and spatial information on proteoforms to reveal how the proteome drives human physiology. The proteomics toolkit offers a variety of methods for detecting protein-protein interactions that range from direct detection, such as affinity-purification mass spectrometry, to less direct, such as correlational profiling or spatial co-localization. These approaches are relatively low throughput, whereas we need high-throughput methods for sensitive analysis of protein-protein interactions in primary tissues. We put forward a reductionist view that enough spatial information—in conjunction with other omics data—can define the state, function, and trajectory of a cell within the tissue microenvironment. Approximately 20 spatial proteomics technologies are currently used within the proteomics field;[54,84-91] these approaches can be targeted and multiplexed (e.g., using antibodies) or non-targeted (e.g., MS). Spatial technologies can measure specific protein-protein interactions, subcellular structures, individual cells, and tissue microenvironments within whole organs.

Imaging MS[85,89,92,93] assesses large numbers of glycans, lipids, and metabolites at spatial resolutions of 10 microns.[85,94,95] At the proteomics level, imaging MS can currently identify more than 2000 proteins at spatial resolutions of 100 micron.[89] Using different proteases and glycosidases, imaging MS also can be integrated with, for example, microscopy of the extra-cellular matrix to evaluate multiple analytes sequentially on the same tissue slice.[73,96] A future challenge will lie in applying these methods to resolve proteoforms down to 1 micron. To do so, major bottlenecks like low proteome coverage (especially for large and low abundance proteins) will need to be overcome.

Other imaging MS approaches we envision would combine artificial intelligence powered image analysis with automated single-cell laser microdissection and ultra-high sensitivity MS to link protein abundance to complex cellular or subcellular phenotypes while preserving spatial context. They have successfully identified rare cell states with distinct morphology.[97] Spatial proteomics approaches are also commonly combined with magnetic resonance/computed tomography imaging data to assess anatomical features, stained microscopy to assess tissue morphology, multiplexed immunofluorescence approaches to differentiate cell types, and fluorescence in situ hybridization to observe DNA- and RNA-mediated influences. For example, multimodal imaging analyses of the murine spleen and pancreas have resolved pixel-level cellular populations and subregions.[95,98]

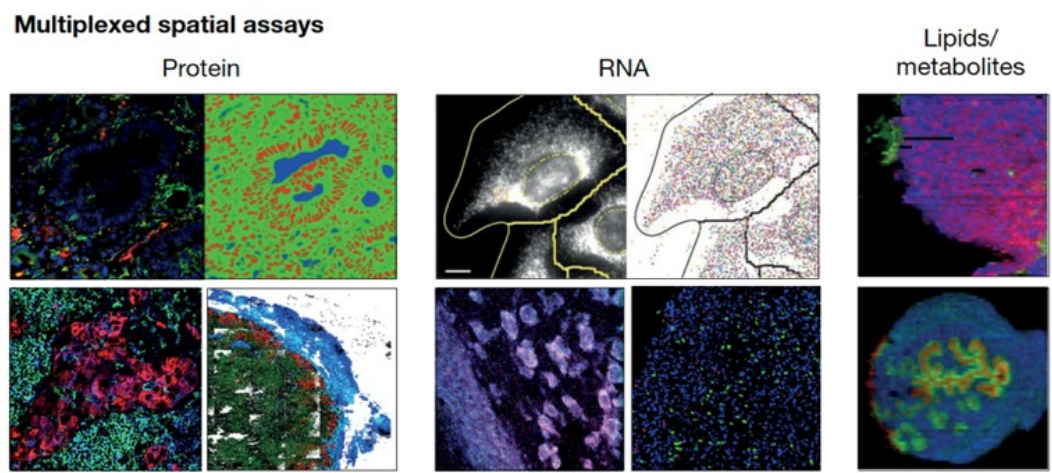

**Figure 4.** Multiplexed spatial assays. Source: The HubMAP Consortium. *Nature*. 2019 Oct 574 (7777):187-192.

Another future challenge with imaging MS lies in tackling sub-cellular resolution. Exploiting fluorescence-based technologies, multiplexed protein imaging can now detail cell subtypes (**Figure 4**). In the future, these protein imaging efforts could also include RNA information. We imagine that the future will bring a new kind of three-dimensional 'AtomScope' capable of measuring every atom in a model cell, then reconstructing single molecules in a bottom-up process. As with single-cell and single-molecule proteomics, such aspirations will benefit from reference atlases and improved baseline knowledge about the composition of the proteome. With meaningful data integration, models of the future will identify spatial dependencies of compositional data, correlate spatial patterns to specific outcomes, assess relationships between data types to enable predictions across them, and integrate data generated across bulk and single cell scale. One early example lies in integration of protein imaging with protein-protein interaction data in the Multi-Scale Integrated Cell model[99] that led to the identification of novel cellular systems.

## 5b. Meaningful data integration

Given the above developments, the way forward is to integrate different data sets in a biologically informed and statistically robust manner. Such integration will then detect emergent properties of the system as a whole,[100] and consider the cell and its proteome as an 'ecosystem' in which the sum is more than the parts. One popular method of integration exploits networks, with mathematical approaches available to understand and predict properties (**Figure 5A**). Networks include both parts lists and interactions among the parts. Constructing such networks will allow us to understand concepts such as the robustness of a system (the cell) to perturbations (i.e., mutations and loss of the protein), or efficiency and redundancy in signal transduction (i.e., through creation of short paths through a network that prioritize rapid response over versatility). Future work will accommodate different types of nodes (e.g., proteoforms) and different types of edges (e.g., kinase-target interactions or protein-protein interactions in a complex).

We propose that in the coming decade, 'filling this network with life' will involve re-thinking commonly used terms (e.g., protein function). While functional descriptions are standardized in a controlled vocabulary such as Gene Ontology,[101] we will need to identify more quantifiable definitions (e.g., by the profile of the genetic, physical, or regulatory interactions of a protein). Such definitions would also allow us to measure similarity and changes to proteoform function. For example, once a transcript is spliced, rearranged, and modified, the resulting different proteoforms are expected to mediate divergent functions, or behave as *alloforms,* rather than retain similar functions, or acting as *isoforms*.[102]

Other proteomics integration efforts will need to venture toward other molecule types, such as RNA, or metabolites that are already being captured with spatial resolution (**Figure 5**). Existing efforts have already shown the power of such an approach to infer regulatory mechanisms and most tightly connected, and therefore potentially most functionally important, relationships. For example, integrating complementary RNA, protein, and metabolite in the Kidney Precision Medicine Project (KPMP) has generated many new hypotheses.[103] More generalized efforts for multi-omics data integration across space and time are exemplified by the Human Tissue Atlas Network (HTAN).[104]

Finally, we propose that the future of proteomics lies not only in integrative network approaches and re-defining entities to render them quantifiable, but will also include consideration of the combinatorial nature of biological processes. Traditionally, we study one protein or pathway at a time as the cell is otherwise too complex. Including combinatorial action of proteoforms under the same conditions, attacking the same problem (e.g., through above-mentioned network approaches) will allow us to truly understand cells and functional tissue units as systems driving life processes such as those shown in the bottom of **Figure 5B**. While constructing networks, quantifying protein activity, defining proteoforms, and considering combinatorial actions are being addressed in specific areas, one future of functional proteomics requires simultaneous implementation of *all* these approaches to develop a comprehensive, predictive framework of cell and whole tissues for the first time.

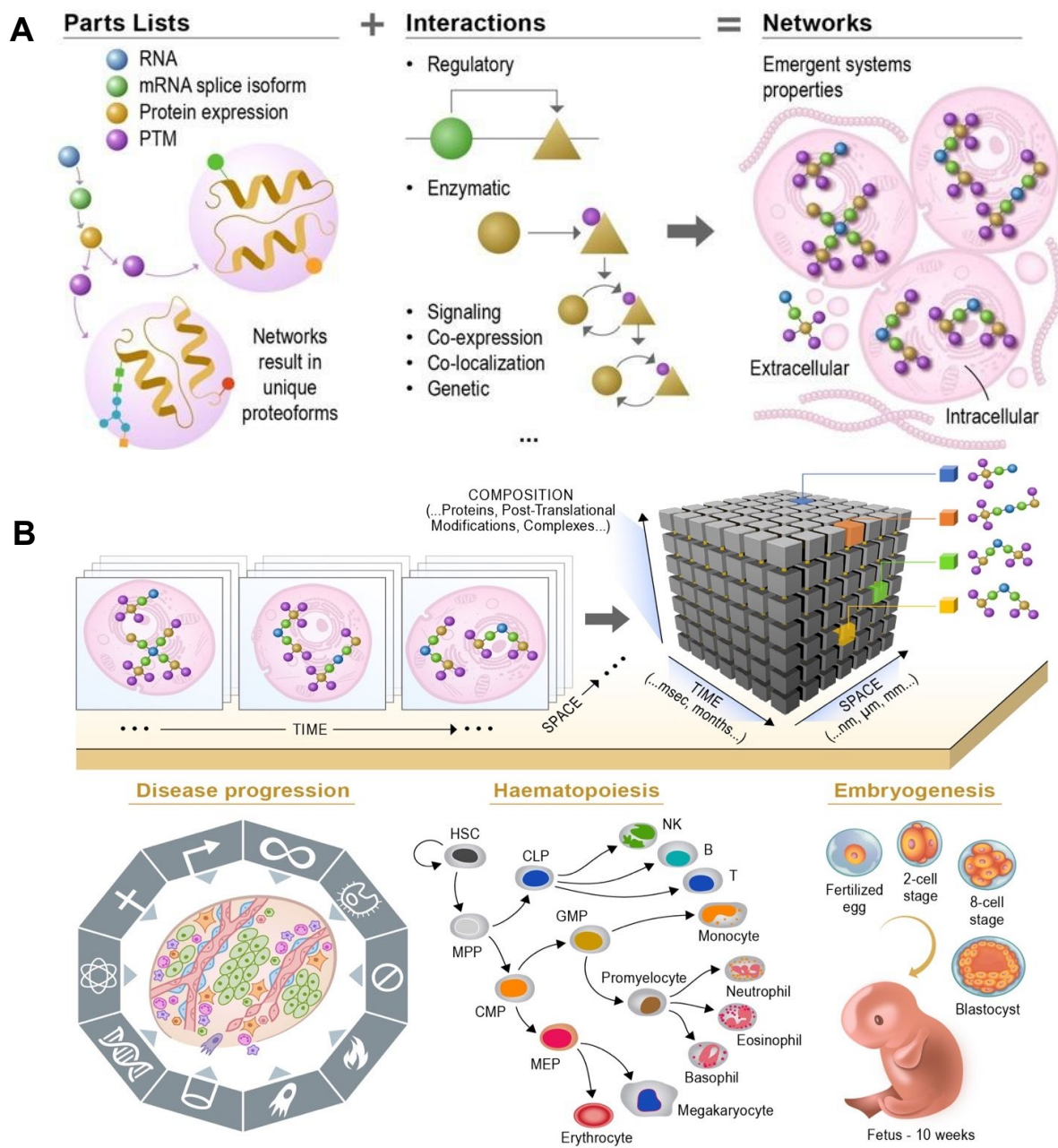

**Figure 5.** *Capturing the biology of our proteins.* A. From a parts lists to interaction mapping and proteoform networks underlying cell-based biology. B. Capturing time-resolved human biology at the protein level. (Top) Depiction of multiple spatio-molecular proteoform images collected over time. (Bottom) Temporal dynamics vary widely across disease progression (years) across cell differentiation (days) in human bone marrow and blood, or through embryogenesis (months).

## Section 6. Meeting the scale of the challenge

Given the vast challenge of capturing protein-level biology (depicted as a *hypercube* in **Figures 5B, 6A**) and the state of measurement, we argue for an increased rate of progress in proteomics. Combinations of new, accessible technologies need to provide synergistic data streams to move us beyond census taking and tenuous inferences to far greater understanding of how protein function and how they go awry. The future of proteomics clearly lies in expanding its 'dimensions' and building stronger connections across the spatial, temporal, and compositional domains. To achieve major outcomes in biomedicine (**Figure 6C**), we should move beyond partial and siloed information on proteins, to an increasingly integrated field[105] providing functional information on a bounded, defined proteome.

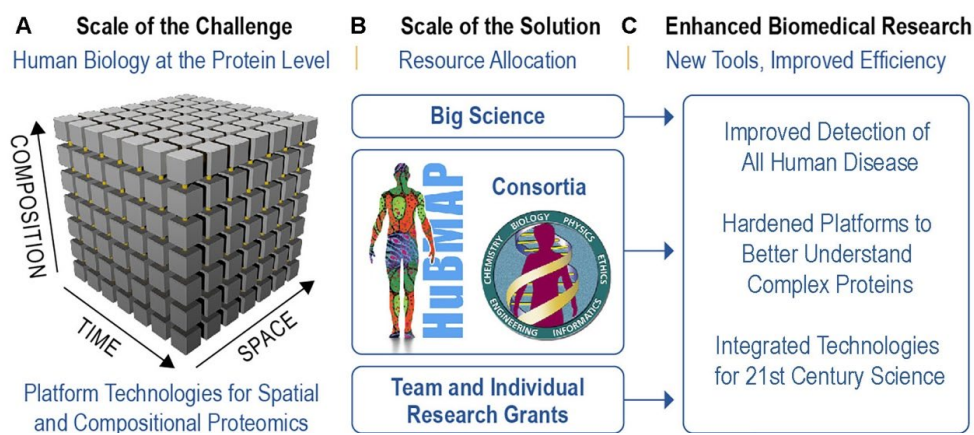

**Figure 6.** *Collaborative and big science models for approaching the human proteome in the decade ahead.* Scope of the solution needs to match scope of the challenge, if we are to make interactions with the human proteome more deterministic for the goals of biomedical research, including regenerative medicine, more efficient drug development and early detection of all types of human disease.

Science is inherently a multi-scale endeavor (**Figure 6B**). In addition to ground-breaking work in individual research laboratories, there are many initiatives building on and maximizing information gain at a larger scale, including the Human Cell Atlas,[106] the Human Proteoform Project (CTDPs),[38] the Human Proteome Project (HUPO),[6] the Human Protein Atlas,[7] and HuBMAP,[8] and many organ- or disease-specific consortia such as KPMP,[107] HTAN,[104] etc. Increasingly, consortia are elevating and integrating proteomics with related fields using finite resources. Working across silos, consortia have demonstrated science explored at a scale not possible in typical single-investigator laboratories, to convene multidisciplinary communities, and build resources in close collaboration with the research community. Currently, even large teams are only able to generate a small portion of the data needed to comprehensively capture the enigmatic proteome (**Figure 6A**). Democratize access to large datasets from adjacent atlasing projects is essential. There is a natural tension between building a multiscale scientific approach and the need to be nimble, implement new technologies, and effectively facilitate training and adoption. However, many consortia are responding with agile management and investigator buy-in to achieve unity-of-engagement and economy-of-force.

All human diseases involve proteins, and there is an exciting range of new tools providing new insights in protein composition, spatial patterning and temporal dynamics. A multitude of technologies and multidisciplinary expertise are required,[105] and extensive interactions between computational, statistical, clinical, and biology experts will be more important than ever. Proteomics needs to be deeply cross-disciplinary, providing the data and connections to start capturing the complexity of the cell, functional tissue units and complex traits of an entire person. To accomplish this in proteomics, we need hardened platforms capable of powering quantitative models to reliably connect protein information to disease and unlock precise decision-making (**Figure 6C**). With well-coordinated efforts along lines outlined here, tremendous advances are possible within a decade, but only with increased level of focus and intention. To advance beyond our current trajectory, a big science, 'proteomic moonshot' will be needed to reach 'criticality' with the human proteome by 2030.

## Acknowledgements

CV was supported by award R35GM127089, 75N93019C00052, and the Chan Zuckerberg Initiative; KBJ was supported by award UH3 CA255132-03; NLK by UH3 CA246635-02; GN by U54 HG010426-03; and JS U54 DK120058-03 all of which are part of the Human Biomolecular Atlas Program (HuBMAP). AN was supported by R01 CA232517 and Catalyst Award C-088 from the Chicago Biomedical Consortium. AEH was supported by a NIH Cancer Moonshot grant (R33CA225296) and as a Chan Zuckerberg Biohub Investigator. EL was supported by the KAW Foundation (2016.0204, 2018.0172), the Swedish Research Council (2017-05327) and the Chan Zuckerberg Initiative (CZF2019-002448).NS was supported by DP2GM123497 and CZF2019-002424, and NS and NLK supported by Allen Distinguished Investigator awards through the Paul G. Allen Frontiers Group. RTK was supported by R01 GM138931. The authors would like to thank Nathan Johnson (Pacific Northwest National Laboratory) and Michael Mullowney for assistance in generating figures. The opinions expressed in this article are the authors' own and do not necessarily reflect the views of the National Institute of Diabetes and Digestive and Kidney Diseases.

## Conflicts of Interest

TPC is a ThermoFisher Scientific Advisory Board member and receives research funding from AbbVie. AEH receives research support from Thermo Fisher Scientific. JVE is a Bruker Software Development SAB member and receives support from ThermoFisher, Sciex, Agilent, CellenONE, and Neoteryx. Many authors are involved with private sector endeavors and actively manage competing financial interests.

**Supplementary Image and Context of NIH-sponsored Workshop:**

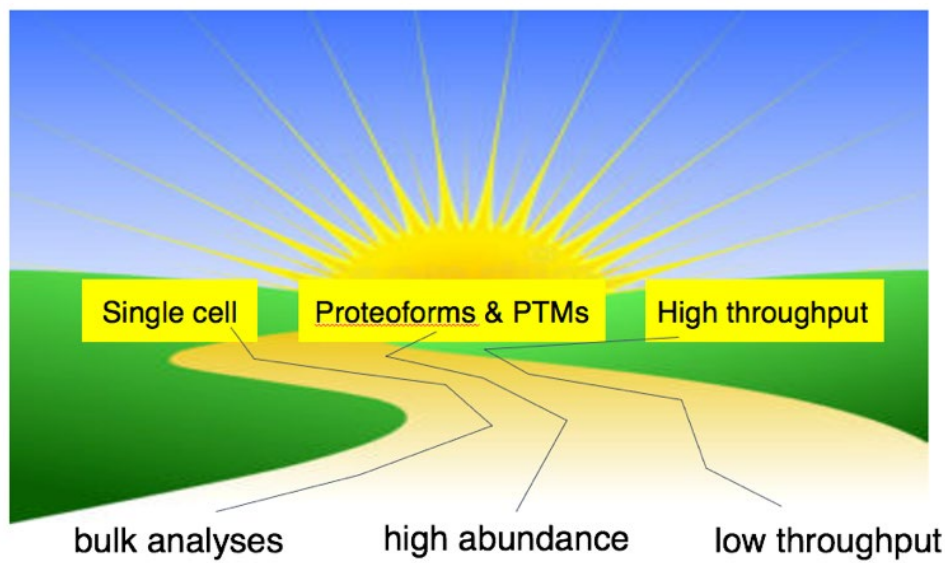

*Background.* On September 8-9, 2020, the National Institutes of Health (NIH) Office of Strategic Coordination (OSC) convened a virtual meeting to foster discussion among experts on existing gaps and opportunities within the functional proteomics field. Organized by the NIH OSC, in collaboration with Workshop Co-Chairs and HuBMAP-funded principal investigators, this meeting engaged proteomics and atlasing experts in discussions designed to 1) to identify how functional proteomics tools, methods, and datasets can be integrated to accelerate the development of comprehensive spatio-molecular tissue maps that generate new biological insights and 2) develop a prospective publication that articulates a vision of the field, including state-of-the-art and near-future tools and methods, a roadmap for integrating multiple proteomics data types (e.g., spatial proteomics, single-cell proteomics), and an articulation of the most important knowledge gaps.

**Corresponding authors**

*Correspondence to Kristin E. Burnum-Johnson (kristin.burnum-johnson@pnnl.gov) or Neil L. Kelleher (n-kelleher@northwestern.edu)

**Co-Authors and Affiliations**

[1]**Kristin E. Burnum-Johnson**
The Environmental Molecular Sciences Laboratory
Pacific Northwest National Laboratory, Richland, Washington, United States
[2]**Thomas P. Conrads**
Inova Women's Service Line,
Inova Health System, Falls Church, Virginia, United States

[3]**Richard R. Drake**
Cell and Molecular Pharmacology and Experimental Therapeutics
Medical University of South Carolina, Charleston, South Carolina, United States
[4]**Amy E. Herr**
Department of Bioengineering
University of California, Berkeley, California, United States
[5]**Ravi Iyengar**
Department of Pharmacology and Systems Therapeutics
Icahn School of Medicine at Mount Sinai, New York, New York, Unites States
[6]**Ryan T. Kelly**
Department of Chemistry and Biochemistry
Brigham Young University, Provo, Utah, United States
[7]**Emma Lundberg**
Science for Life Laboratory
KTH Royal Institute of Technology, Stockholm, Sweden
[8]**Michael J. MacCoss**
Department of Genome Sciences
University of Washington, Seattle, Washington, United States
[9]**Alexandra Naba**
Department of Physiology and Biophysics
University of Illinois at Chicago, Chicago, Illinois, United States
[10]**Garry P. Nolan**
Dept of Pathology
Stanford University, Stanford, California, United States
[11]**Pavel Pevzner**
Department of Computer Science and Engineering
University of California at San Diego, San Diego, California, United States
[12]**Karin D. Rodland**
Biological Sciences Division
Pacific Northwest National Laboratory, Richland, Washington, United States
[13]**Salvatore Sechi**
National Institute of Diabetes and Digestive and Kidney Diseases,
National Institutes of Health, Bethesda, Maryland, United States
[14]**Nikolai Slavov**
Department of Bioengineering
Northeastern University, Boston Massachusetts, United States
[15]**Jeffrey M. Spraggins**
Department of Chemistry
Vanderbilt University, Nashville, Tennessee, United States
[16]**Jennifer E. Van Eyk**
Advanced Clinical Biosystems Institute in the Department of Biomedical Sciences
Cedars-Sinai Medical Center, Los Angeles, California, United States
[17]**Marc Vidal**
Department of Genetics
Harvard University, Cambridge, Massachusetts, United States

[18]**Christine Vogel**
New York University Center for Genomics and Systems Biology
New York University, New York, New York, United States
[19]**David R. Walt**
Department of Pathology
Harvard Medical School, Brigham and Women's Hospital, Wyss Institute at Harvard
University, Boston, Massachusetts, United States
[20]**Neil L.  Kelleher**
Department of Chemistry
Northwestern University, Evanston, Illinois, United States